\documentclass[showpacs,pra,twocolumn,amsmath,amssymb]{revtex4-1} %

\usepackage{epsfig,amsmath}
\usepackage{subfigure}
\usepackage{graphicx}
\usepackage{dcolumn}
\usepackage{stmaryrd}
\usepackage{mathrsfs}
\usepackage{pifont}
\usepackage{amsthm}
\usepackage{amssymb}
\usepackage{bm}
\usepackage{latexsym}
\usepackage{hyperref}

\newcommand{\beq}{\begin{equation}}
\newcommand{\eeq}{\end{equation}}
\newcommand{\beqa}{\begin{eqnarray}}
\newcommand{\eeqa}{\end{eqnarray}}

\newcommand{\pa}{\partial}
\newcommand{\non}{\nonumber}
\newcommand{\la}{\langle}
\newcommand{\ra}{\rangle}
\newcommand{\ti}{\tilde}
\newcommand{\da}{\dagger}

\newcommand{\lam}{\lambda}
\newcommand{\ga}{\gamma}
\newcommand{\Ga}{\Gamma}
\newcommand{\ka}{\kappa}
\newcommand{\De}{\Delta}

\newcommand{\om}{\omega}

\def\jpb#1{{ J.\ Phys.\ B} {\bf#1}}

\def\pra#1{{ Phys.\ Rev. A\/} {\bf#1}}
\def\prb#1{{ Phys.\ Rev. B\/} {\bf#1}}
\def\pre#1{{ Phys.\ Rev. E\/} {\bf#1}}
\def\prl#1{{ Phys.\ Rev.\ Lett.} {\bf#1}}

\def\pla#1{{ Phys.\ Lett. A\/} {\bf#1}}

\def\nat#1{{ Nature} {\bf#1}}

\begin{document}

\title{Feshbach Projection Operator Partitioning for Quantum Open Systems:
Stochastic Approach} 

\author{Jun Jing$^{1}$\footnote{Email address: Jun.Jing@stevens.edu},
Lian-Ao Wu$^{2,3}$\footnote{Email address: lianao_wu@ehu.es},
J. Q. You$^{4}$\footnote{Email address: jqyou@fudan.edu.cn},
Ting Yu$^{1}$\footnote{Email address: Ting.Yu@stevens.edu}}

\affiliation{$^{1}$Center for Controlled Quantum Systems and Department of
Physics and Engineering Physics, Stevens Institute of Technology, Hoboken, New
Jersey 07030, USA \\ $^{2}$Ikerbasque, Basque Foundation for Science, 48011
Bilbao, Spain \\ $^{3}$Department of Theoretical Physics and History of
Science, The Basque Country University (EHU/UPV), PO Box 644, 48080 Bilbao,
Spain \\ $^{4}$Department of Physics and State Key Laboratory of Surface
Physics, Fudan University, Shanghai 200433, China}

\date{\today}

\begin{abstract}
Dynamics of a state of interest coupled to a non-Markovian environment is
studied for the first time by concatenating the non-Markovian quantum state 
diffusion (QSD) equation and the Feshbach projection operator partitioning
technique. An exact one-dimensional stochastic master equation is derived as a
general tool for controlling an arbitrary component of the system. We show that
the exact one-dimensional stochastic master equation can be efficiently solved
beyond the widely adapted second-order master equations. The generality and
applicability of this hybrid approach is justified and exemplified by several
coherence control problems concerning quantum state protection against leakage
and decoherence.
\end{abstract}


\maketitle

\section{Introduction}

Feshbach projection operator partitioning technique (termed PQ partitioning for
brevity) allows one to focus exclusively on the dynamics of a small subspace in
the whole Hilbert space of an open or closed quantum system
\cite{Gaspard,Wu09}. This has been proven to be exceptionally useful
in dealing with the decoherence suppression for quantum states within many
interesting contexts in physics such as quantum state storage where PQ
partitioning can significantly reduce the resource by focusing on a target
state rather than on the entire multi-level (even infinite level) Hilbert
space. The targeted component in the Hilbert space, called P subspace, could be
a superposition of some or all of the energy levels, while the rest of the
state space is denoted as Q subspace. Generally there is a bidirectional
wave-function flow between P and Q subspaces, unless some intervention
mechanism is implemented. A typical example of this case is the Quantum Zeno
Effect, that may decompose the whole space of the system into isolated Zeno
subspaces \cite{Kofman,QZE} and the remaining parts of the whole space. When
the system is embedded in a dissipative environment, the dissipation of the
whole system and the leakage of the subspace will mix with each other to leave
a compelling, yet subtle question of the dynamical control of one subspace in
open quantum systems, especially in those with a non-Markovian environment.

A majority of prior research efforts on a non-Markovian environment are based
on a second-order master equation in terms of coupling constant obtained by
using projection-operator technique \cite{Gordon} or the Liouvillean approach
\cite{Puri}, where the weak coupling regime is typically assumed. The work
presented in this paper will investigate the dynamics and control of an
arbitrary subspace of an open system beyond the weak non-Markovian regime by
employing the non-Markovian quantum state diffusion equation
\cite{Diosi1,Diosi2,Strunz,YDGS99,Jing}, which is capable of dealing with the
strong coupling strength and the arbitrary correlation function of the
environment. The non-Markovian QSD equation may be cast into a
convolutionless form, hence it also serves as a useful tool in deriving the
corresponding exact master equation \cite{YDGS99,Strunz-Yu2004,Yu2004}. In
addition, the approximate QSD equation obtained by perturbation may also
include the contributions from the high-order non-Markovian master equation
\cite{Breuer,Maniscalco,Goan}.

Based on a microscopic quantum dissipative model, we will first derive a
general one-dimensional dynamical equation for a subspace by combining the QSD
equation and the PQ partitioning technique. The system survival probability in
the subspace is obtained by the ensemble average over the trajectories for the
survival amplitude. For simplicity but without loss of generality, we consider
an initial pure state as our subspace of interest (with dimension one). 

We show in this paper via the stochastic approach that the non-Markovian
dynamics of the one dimensional subspace can be effectively controlled by a
sequence of pulses applied to the system. In particularly, we show that the
control function $c(t)$ can be chosen as a simple periodic rectangular
interaction \cite{Zhang}: $c(t)=\frac{\Psi}{\De}$, where $\Psi$ is the
interaction intensity, for regions $n\tau-\De<t<n\tau$, $n\geqslant1$ integral;
otherwise $c(t)=0$, where $1/\tau$ is the frequency of the pulse and $\De$ is
the duration time of the pulse in one period. The pulses applied here are
neither an ideal zero-width pulse nor an optimized Bang-Bang control
\cite{Gong}. Yet it is sufficient for our control scheme if the interval $\tau$
is short enough. We find that increasing $\Psi$ is also beneficial to reduce
the leakage rate of the subspace.

\section{Feshbach PQ partitioning and one-dimensional stochastic master
equation} 

We consider an open quantum system with $(N+1)$ normalized base vectors ($N$ is
arbitrary) coupled to a bath of harmonic oscillators described by the following
total Hamiltonian (setting $\hbar=1$): 
\beq \label{tot}
H_{\rm tot}=H_{\rm sys}+\sum_\lam(g_\lam^*La_\lam^\da+g_\lam L^\da a_\lam)
+\sum_\lam \om_\lam a_\lam^\da a_\lam,
\eeq
where $H_{\rm sys}$ and $L$ are the system Hamiltonian and the Lindblad
operator, respectively. The system state described by a stochastic
wave-function $\psi_t$ is governed by the non-Markovian QSD equation
\cite{Diosi1,Diosi2}:
\beq \label{1QSD}
i \partial_t \psi_t 
= \left[H_{\rm sys}+iLz_t^*-iL^\da\bar{O}(t,z^*)\right]\psi_t
= H_{\rm eff}\psi_t.
\eeq
Here $z^*_t$ is the colored noise arising from coupling to the environment such
that its statistical mean recovers the environment correlation function
$\alpha(t,s)$: $M[z_t z_s^*]=\alpha(t,s)$. Note that $\bar{O}(t,z^*)$, with its
explicit expression given below, is the system operator representing the effect
of the environment. Thus the effective Hamiltonian $H_{\rm eff}$ contains all
the information about the open system and its interaction with the environment.
Whenever the operator $\bar{O}$ is exactly constructed, then effective
Hamiltonian is exact. That is, it is directly derived from the total
Hamiltonian without using any approximations, in particular, without 
Born-Markov approximation.

The PQ partitioning technique can divide the system wave-function $\psi_t$
(with $(N+1)$-dimension) into two parts: a scalar function $P(t)$ associated
with a chosen vector denoted by $|0\ra$ and an $N$-dimensional vector $Q(t)$.
Note that $|0\ra$ can be an arbitrary component of the system and doest not
necessarily denote the ground state. With this partition, the state and the
effective Hamiltonian may be written as
\begin{equation}\label{Heff}
\psi_t=\left[\begin{array}{c} P \\ \hline Q \end{array}\right], \quad
H_{\rm eff}=\left(\begin{array}{c|c}
      h & R \\ \hline
      W & D
    \end{array}\right).
\end{equation}
Here the $1\times1$-matrix $h$ and the $N\times N$-matrix $D$ correspond to the
self-Hamiltonians living in the P subspace and the Q subspace, respectively.
For the effective Hamiltonian, the resulting $W$ and $R$ are not mutually
conjugate to each other in general. Consequently, the QSD equation (\ref{1QSD})
may be decomposed into two parts:
\begin{eqnarray}
\label{PE} i\partial_tP&=&hP+RQ, \\
\label{QE} i\partial_tQ&=&WP+DQ.
\end{eqnarray}
The solution for Eq.~(\ref{QE}) could be formally expressed by
\begin{equation}\label{Qt}
Q(t)=-i\int_0^tdsG(t,s)W(s)P(s)+G(t,0)Q(0),
\end{equation}
where $\partial_tG(t,s)=-iD(t)G(t,s)$ and $G(t,t)=1$. Note that the propagator
may be written as
\begin{equation}\label{Gts}
G(t,s)=\mathcal{T}_{\leftarrow}\left\{\exp\left[-i\int_s^tD(s')ds'\right]
\right\},
\end{equation}
where $\mathcal{T}_{\leftarrow}$ is the time-ordering operator. Substituting
Eq.~(\ref{Qt}) into Eq.~(\ref{PE}), we obtain a closed one-dimensional
master equation for $P(t)$:
\begin{eqnarray}\non
i\partial_tP(t)&=&h(t)P(t)-i\int_0^tds\ti{G}(t,s)P(s)\\ \non
&+&R(t)G(t,0)Q(0), \\
\ti{G}(t,s)&=&R(t)G(t,s)W(s).\label{ME1}
\end{eqnarray}
In case that matrix $D$ is diagonalizable or
$G(t,s)\approx\sum_{n=1}^Ne^{-i\int_s^tD_{nn}(s')ds'}|n\ra\la n|$ is a good
approximation, Eq.~(\ref{ME1}) can be readily calculated. Especially when
$\ti{G}(t,s)=0$, a formal solution can be given,
\begin{equation}\label{Pt2}
P(t)=\bigg[P(0)-i\int_0^tds'R(s')Q(s')e^{i\int_0^{s'}dsh(s)}
\bigg]e^{-i\int_0^tdsh(s)}.
\end{equation}
Notably, the off-diagonal term $R(t)$ between $P$ and $Q$ plays a vital role in
the dynamics of $P(t)$ while the local control term $h(t)$ only provides a
phase factor.

Equation (\ref{ME1}) will be the main result in the concatenation of the PQ 
partitioning and the non-Markovian QSD equation described by $H_{\rm eff}$. As
shown below, the one-dimensional master equation (\ref{ME1}) can yield some
very interesting analytical results that are hard to obtain without invoking 
certain approximations such as weak-coupling approximation. We will now discuss
three examples, where $\bar{O}(t,z^*)$ is exact, hence $H_{\rm eff}$ is exact
and time-local, to illustrate both the free and controlled dynamics of states
of interest coupled to a zero temperature heat bath, in terms of their fidelity
with respect to the initial state. For an arbitrary pure initial state
$\psi_0$, it is easily to show that the fidelity
${\mathcal F}(t)=M[|P(0)^*P(t)+Q^\dag(0)Q(t)|^2]$. A simpler expression for the
fidelity is obtained ${\mathcal F}(t)=M[|P(t)|^2]$ when
$Q(0)={\bf 0}$, which is equivalent to the population or survival probability
over $\psi_0$. Remarkably, in many physically interesting examples as shown
below, the PQ partitioning combined with the QSD equation can yield simple
analytical solutions of the fidelity ${\mathcal F}(t)$.

\subsection{Dissipative two-level atom model}

\begin{figure}[htbp]
\centering
\includegraphics[width=3.2in]{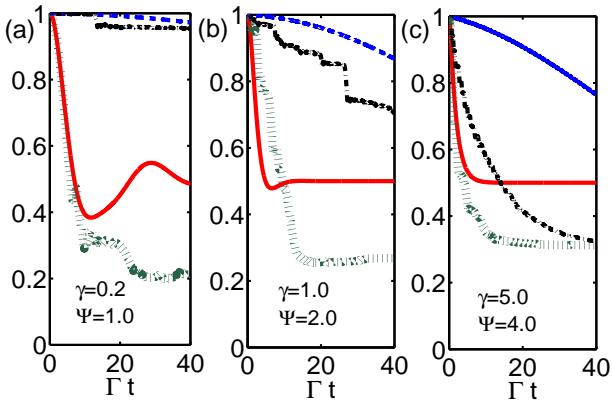}
\caption{Fidelity dynamics of the two-level system with different $\gamma$ and
$\Psi$. Red solid lines for free dynamics, blue dashed lines for
$\tau=2\Delta$, black dot-dashed lines for $\tau=3\Delta$, and green dotted
lines for $\tau=6\Delta$. The other parameters are set as $\om=0.2\Gamma$, and
$\Delta=0.04\Gamma t$.}
\label{two}
\end{figure}

The model is represented by $H_{\rm sys}=E_0(t)|0\ra\la0|+E_1(t)|1\ra\la1|$ and
$L=|0\ra\la1|$. The effective Hamiltonian in the basis $|0\ra, |1\ra$ becomes
\begin{equation}
H_{\rm eff}=\left(\begin{array}{c|c}
      E_0(t) & iz_t^* \\ \hline
      0 & E_1(t)-iF(t)
    \end{array}\right),
\end{equation}
where $F(t)$ is the coefficient function in the operator
$\bar{O}(t,z^*)=F(t)L$ with $F(t)\equiv\int_0^tds\alpha(t,s)f(t,s)$ and the
initial condition $F(0)=0$. The equation of motion for $f(t,s)$ is given by
$\pa_tf(t,s)=[iE(t)+F(t)]f(t,s)$ \cite{Diosi2}, where $E(t)=E_1(t)-E_0(t)$. For
this two-level atom and the systems in the following two examples, we have
$E(t)=\om+c(t)$, where $\om$ is the bare frequency for the system and $c(t)$ is
the control function.

With the initial state of the system $|\psi_0\ra=(1/\sqrt{2})(|0\ra+|1\ra)$,
our aim is to control the population of this chosen state. The exact stochastic
transition amplitude is
\begin{eqnarray}\non
\la\psi_0|\psi_t\ra&=&\frac{1}{2}\bigg\{e^{-i\int_0^tdsE_1'(s)}+
e^{-i\int_0^tdsE_0(s)}\\ &\times& \bigg[1+
\int_0^tds'z_{s'}^*e^{-i\int_0^{s'}ds[E_1'(s)-E_0(s)]}\bigg]\bigg\},
\end{eqnarray}
where $E_1'(s)\equiv E_1(s)-iF(s)$. Consequently in the rotating picture of
$H_{\rm sys}$, we have
\begin{eqnarray}\non
{\mathcal F}(t)&=&\frac{1}{4}\bigg[1+\bar{F}_R^2(t)+
2\bar{F}_R(t)\bar{F}_I(t)+\int_0^t\int_0^tds_1ds_2 \\
\label{tlMP} &\times& \alpha(s_1,s_2)\bar{F}_R(s_1)\bar{F}_R(s_2)
\bar{K}(s_1)/\bar{K}(s_2)\bigg],
\end{eqnarray}
where $\bar{F}_R(t)\equiv e^{-\int_0^tdsF_R(s)}$,
$\bar{F}_I(t)\equiv\cos[\int_0^tdsF_I(s)]$ ($F_R$ and $F_I$ stand for the real
and imaginary parts of $F$ respectively) and
$\bar{K}(t)\equiv e^{i\int_0^tds[E(s)+F_I(s)]}$.

For simplicity, throughout the paper the non-Markovian environmental noise is
described by an Ornstein-Uhlenbeck type correlation \cite{OU}
$\alpha(t,s)=\frac{\Gamma\gamma}{2}e^{-\gamma|t-s|}$, where $\Gamma$ is the
environmental dissipation rate and $1/\gamma$ characterizes the memory time of
the environment. The finite, but non-zero $1/\gamma$ gives a non-Markovian
process with the Markov limit when $\gamma \rightarrow \infty$. The equation of
motion for $F(t)$ can be easily derived from the QSD equation:
\begin{equation}\label{Ft}
\partial_tF(t)=\frac{\Gamma\gamma}{2}+[-\gamma+iE(t)]F(t)+F^2(t).
\end{equation}
Therefore, Eq.~(\ref{tlMP}) may be controled by the functions $E(t)$ (or
$c(t)$) via Eq.~(\ref{Ft}).

When $c(t)=0$, the two-level atom will be driven by the environment into the
ground state after a period of time. The red lines in Fig.\ref{two} show that
the fidelity approaches $0.5$ regardless of the memory times. Clearly, in the
absence of control function, the smaller $\gamma$ typically gives rise
to more significant fluctuations, yet it does not  effectively preserve the
fidelity. However, when the control pulses are applied, we see that the
non-Markovian features can greatly enhance the efficiency of the rectangular
pulses for controlling the state of interest. The blue dashed line in
Fig.\ref{two}(a) indicates that the fidelity can be preserved for a long period
of time. Particularly, when the period of the pulse $\tau$ is chosen as small
as twice the interaction duration time $\Delta$ in one period, the fidelity is
fully preserved up to the time $\Gamma t=13$. Moreover, we show that the
control pulses can work very well even if $\tau$ is as long as the three times
of $\Delta$ (the black dot-dashed line). But if the period of pulses is too
long, the dynamical decoupling becomes highly inefficient \cite{Biercuk}. For
example, when $\tau=6\Delta$, as shown by the green dotted lines in
Figs.\ref{two}(a), \ref{two}(b) and \ref{two}(c), the control function $c(t)$
becomes adversary of the fidelity control. In all those three cases, the
fidelity decays faster than those of free evolution ({\it i.e.}, $c(t)=0$).
With decreasing memory time of the environment plotted in Figs.\ref{two}(b) and
\ref{two}(c), the control effect of the fidelity is gradually weakened even
using the same rapid pulses as in Fig.\ref{two}(a) and simultaneously raising
the interaction intensity $\Psi$ for partial compensation. When $\gamma=2.0$
and $\tau\geqslant3\Delta$, we see that the control becomes
inefficient in the Markov limit (large $\gamma$).

\subsection{A qutrit dissipative model}

\begin{figure}[htbp]
\centering
\includegraphics[width=3.2in]{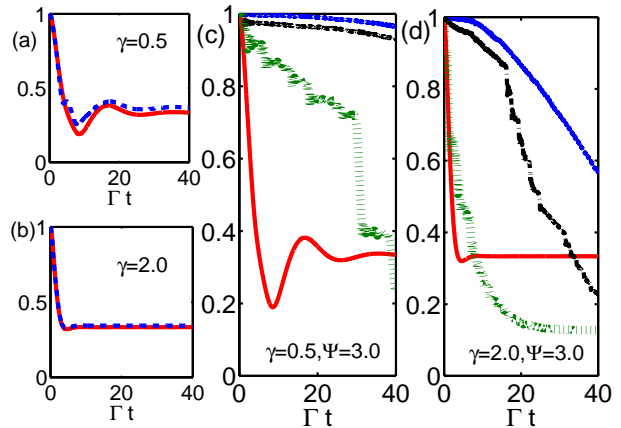}
\caption{Fidelity dynamics of the qutrit system with different $\gamma$.
In (a) and (b), the approximate results (blue dashed lines) by
Eq.~(\ref{appro}) is compared with the exact ones (red solid lines); In (c) and
(d), Red solid lines for free dynamics, blue dashed lines for $\tau=2\Delta$,
black dot-dashed lines for $\tau=3\Delta$, and green dotted lines for
$\tau=6\Delta$. The other parameters are set as $\om=1.0\Gamma$,
$\Delta=0.04\Gamma t$, and $\ka=\sqrt{2}$.}
\label{three}
\end{figure}

For this model, we have $H_{\rm sys}=E(t)(|2\ra\la2|-|0\ra\la0|)$ and
$L=\ka(|0\ra\la1|+|1\ra\la2|)$. In the basis $|1\ra$, $|0\ra$ and $|2\ra$, the
effective Hamiltonian takes for the following form:
\begin{equation}\label{appro}
H_{\rm eff}=\left(\begin{array}{c|cc}
      -iF_1 & 0 & i\ka z_t^*-i\ka U_z \\ \hline
      i\ka z_t^* & -E & 0 \\
      0 & 0 & E-iF_2
    \end{array}\right),
\end{equation}
where $F_1(t)$, $F_2(t)$ and $U_z(t)=\int_0^tds'U(t,s')z_{s'}^*$ are the three
coefficient functions in
$\bar{O}(t,z^*)=F_1(t)|0\ra\la1|+F_2(t)|1\ra\la2|+U_z(t)|0\ra\la2|$. One
particularly interesting feature of this model is that in the PQ partitioning
$W \neq 0$, but $D$ is diagonal, so it allows a simple solution. Applying the
noise-free approximation, we have
$\bar{O}(t,z^*)=F_1(t)|0\ra\la1|+F_2(t)|1\ra\la2|$ with
\begin{eqnarray}\non
\frac{dF_1(t)}{dt}&=&\frac{\Ga\ga\ka^2}{2}+(-\ga+iE)F_1+F_1^2-F_1F_2, \\
\frac{dF_2(t)}{dt}&=&\frac{\Ga\ga\ka^2}{2}+(-\ga+iE)F_2+F_2^2,
\end{eqnarray}
and $F_j(0)=0$, $j=1,2$. The approximation validity in Eq.~(\ref{appro}) is
testified by Fig.\ref{three}(a) and Fig.\ref{three}(b) with two different
$\gamma$'s. We could also find the analytical expression of the fidelity by
PQ partitioning technique. When $|\psi_0\ra=(1/\sqrt{3})(|0\ra+|1\ra+|2\ra)$,
the compact formula for the fidelity is given by
\begin{eqnarray}\non
{\mathcal F}(t)&=&\frac{1}{9}\bigg\{|1+\bar{F}_1(t)+\bar{F}_2(t)|^2+\int_0^tds
\int_0^tds_1\alpha(s_1,s) \non \\
&\times&[\bar{F}^*_1(t)\bar{B}^*(s_1)+\bar{F}^*_1(s_1)\bar{E}^*(s_1)]
[\bar{F}^*_1(t)\bar{B}^*(s_1)\non \\ &+&\bar{F}^*_1(s_1)\bar{E}^*(s_1)]
+\int_0^tds\int_0^{s}ds'\int_0^tds_1\int_0^{s_1}ds_2 \non \\ &\times&
[\alpha(s_1,s)\alpha(s_2,s')
+\alpha(s_1,s')\alpha(s_2,s)] \non \\ &\times&
\bar{F}^*_1(s_1)\bar{E}^*(s_1)\bar{B}^*(s_2)
\bar{F}_1(s)\bar{E}(s)\bar{B}(s')\bigg\},
\end{eqnarray}
where $\bar{F}_j(t)\equiv e^{-\int_0^tdsF_j(s)}$, $j=1,2$,
$\bar{E}(t)\equiv e^{-i\int_0^{t}dsE(s)}$, and
$\bar{B}(t)\equiv \bar{E}(t)\bar{F}_2(t)/\bar{F}_1(t)$.

In Figs.\ref{three}(c) and \ref{three}(d), we compare the control dynamics with
the same environments as in Figs.\ref{three}(a) and \ref{three}(b)
respectively. As in the first example, we also notice the larger environmental
memory time (less $\gamma$) is very helpful to relieve the requirement of the
pulse frequency than the smaller one (bigger $\gamma$). If $\gamma=0.5$,
$\tau\leqslant3\Delta$, the fidelity could be maintained above $0.9$ even when
$\Gamma t$ approaches 40. In Fig.\ref{three}(d), $\gamma=2.0$, it is
interesting to observe a damping enhancement of the fidelity especially when
$\tau\geqslant3\De$. This (as well as all the green dotted lines in
Fig.\ref{two}) corresponds to a sort of Quantum Anti-Zeno Effect (AZE)
\cite{aZeno}. AZE occurs when the evolution is repetitively interrupted by
projecting the system onto some state as in a measurement process \cite{Saha},
by periodically applying the pulses \cite{Kurizki} with long time intervals, or
by coupling between the excited state and an auxiliary state \cite{Sun}. Here
it is induced by unitary pulse sequences, which renormalize the frequency of
the system with a proper period and mimic a repeated measurement process.

\subsection{A special $(N+1)$-level atom}

\begin{figure}[htbp]
\centering
\includegraphics[width=3.2in]{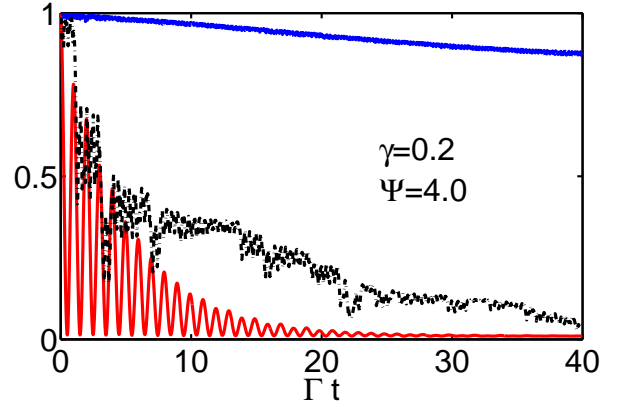}
\caption{Fidelity dynamics of a $101$-level system in a non-Markovian
environment. Red solid line for free dynamics, blue dashed line for
$\tau=2\Delta$, and black dot-dashed lines for $\tau=3\Delta$. The other
parameters are set as $\om=0.2\Gamma$, $\Delta=0.04\Gamma t$.}
\label{np1}
\end{figure}

For this special case, we consider a genuine multi-level atomic system
represented by $H_{\rm sys}=\sum_{n=0}^NE_n|n\ra\la n|$ and
$L=\sum_{n=1}^N|0\ra\la n|$, where $E_0=-E(t)$ and $E_{n\neq0}=E(t)$, and for
this model $\bar{O}(t,z^*)=F(t)L$. Therefore,
\begin{equation}
H_{\rm eff}=\left(\begin{array}{c|cccc}
      -E & iz_t^* & iz_t^* & \cdots & iz_t^* \\ \hline
      0 & E-iF & -iF & \cdots & -iF \\
      0 & -iF  & E-iF & \cdots & -iF \\
      \cdots & \cdots & \cdots & \cdots & \cdots \\
      0 & -iF & -iF & \cdots & E-iF
    \end{array}\right),
\end{equation}
where $F(t)$ satisfies $\pa_tF(t)=\frac{\Ga\ga}{2}+(-\ga+2iE)F+NF^2$ and
$F(0)=0$. Due to the time symmetry of $D$, $[D(t),D(s)]=0$ ($t\neq s$), the
propagator in Eq.~(\ref{Gts}) can be exactly obtained as
\begin{eqnarray}\non
G(t,0)&=&\frac{1}{N}\bigg\{\sum_{j=1}^N\left[\bar{E}(t)\bar{F}^{N}(t)+
(N-1)\bar{E}(t)\right]|j\ra\la j|\\ &+&
\sum_{n\neq m}[\bar{E}(t)\bar{F}^{N}(t)-\bar{E}(t)]|n\ra\la m|\bigg\},
\end{eqnarray}
where $\bar{F}(t)\equiv e^{-\int_0^tF(s)ds}$. For the initial state
$|\psi_0\ra=(1/\sqrt{N+1})\sum_{n=0}^N|n\ra$, again the PQ partitioning allows
to find an analytical expression for the fidelity function:
\begin{align}\non
{\mathcal F}(t)=\frac{1}{(N+1)^2}\bigg[1+N^2\bar{F}_R^{2N}(t)+
2N\bar{F}_R^{N}(t)\bar{F}'_I(t)+N^2 \\ \label{mlMP}
\times\int_0^t\int_0^tds_1ds_2\alpha(s_1,s_2)\bar{F}_R^N(s_1)\bar{F}_R^N(s_2)
\bar{K}'(s_1)/\bar{K}'(s_2)\bigg],
\end{align}
where $\bar{F}'_I(t)\equiv\cos[\int_0^tdsNI_F(s)]$ and
$\bar{K}'(t)\equiv e^{i\int_0^tds[2E(s)+NI_F(s)]}$. It is easy to check that
when $N=1$, Eq.~(\ref{mlMP}) reduces to Eq.~(\ref{tlMP}). It is interesting to
consider the weak coupling limit when $F\rightarrow0$, then $D$ becomes a
diagonal matrix ${\rm diag}[E,E,\cdots,E]$, thus Eq.~(\ref{Gts}) can be
simplified as \begin{eqnarray}\non
{\mathcal F}(t)&=&\frac{1}{(N+1)^2}\bigg[1+N^2+2N+N^2\int_0^t\int_0^tds_1ds_2
\\ &\times& \alpha(s_1,s_2)\bar{E}^2(s_2)/\bar{E}^2(s_1)\bigg].
\end{eqnarray}

We emphasize that our formalism works for an arbitrary $N+1$ system. For
example, we take $N=100$ shown in Fig.\ref{np1}. Without the control function
$c(t)$, the fidelity will quickly decays to $1/(N+1)$. Clearly, the efficient
control can be made possible only when the system is far from Markov regime. As
shown in this example, our pulse control scheme works well for the
non-Markovian environment with $\gamma=0.2$. In fact, by setting $\Psi=4.0$,
and $\tau=2\Delta$, we show that ${\mathcal F}(\Gamma t=40)$ can be maintained
as high as $0.85$.

\section{Conclusion}

We have derived an exact one-dimensional stochastic master equation based on
the Feshbach PQ partitioning and the non-Markovian QSD equation and apply it to
the quantum control dynamics of three distinct model systems. The periodical
rectangular pulses are used to protect the subspace of interest from leakage to
the other part of the Hilbert space of the system and the environment. The
control dynamics measured by time-dependent fidelity is realized by tuning the
pulse frequency $1/\tau$ with different memory time $1/\ga$ of the
non-Markovian environment. Our results have exemplified the simplicity and
power of the exact one-dimensional stochastic master equation. Our approach has
paved a way for studying the direct control dynamics of an arbitrary
multi-level atomic system without invoking the exact or perturbative
non-Markovian master equations. Our hybrid technique would be versatile enough
to accommodate other types of quantum control if some necessary modifications
on the non-Markovian QSD equations are made. It is also possible to consider
the quantum control dynamics of multi-particle system \cite{epl} based on the
PQ partitioning. We leave these open questions for the future investigations.

\begin{acknowledgements}
We acknowledge grant support from the NSF No.~PHY-0925174, DOD/AF/AFOSR No.~FA9550-12-1-0001, and the NSFC
Nos.~91121015 and 11175110, a Ikerbasque Foundation Startup, the Basque
Government (grant IT472-10) and the Spanish MEC (Project No.
FIS2009-12773-C02-02).
\end{acknowledgements}

\end{document}